\documentclass[sigconf]{acmart}

\usepackage{enumitem}
\usepackage{xcolor}

\usepackage{acronym}
\usepackage{array}
\usepackage{mdwmath}
\usepackage{mdwtab}

\usepackage{amssymb}
\usepackage{eqparbox}
\usepackage{amsmath}
\usepackage{tablefootnote}

\usepackage{algorithm}
\usepackage{algorithmicx}
\usepackage{algpseudocode}
\usepackage{amsmath}
\usepackage{diagbox}

\floatname{algorithm}{Algorithm}  

\AtBeginDocument{%
  \providecommand\BibTeX{{%
    \normalfont B\kern-0.5em{\scshape i\kern-0.25em b}\kern-0.8em\TeX}}}

\copyrightyear{2022}
\acmYear{2022}
\setcopyright{acmcopyright}
\acmConference[CIKM '22]{Proceedings of the 31st ACM International Conference on Information and Knowledge Management}{October 17--21, 2022}{Atlanta, GA, USA}
\acmBooktitle{Proceedings of the 31st ACM International Conference on Information
and Knowledge Management (CIKM '22), October 17--21, 2022, Atlanta, GA, USA}
\acmPrice{15.00}
\acmDOI{10.1145/3511808.3557290}
\acmISBN{978-1-4503-9236-5/22/10}
\begin{document}

\title{Dismantling Complex Networks by a Neural Model Trained from Tiny Networks}

\author{Jiazheng Zhang}
\affiliation{%
  \institution{Huazhong University of Science and Technology (HUST)}
  \city{Wuhan}
  \country{China}
}
\email{jiazhengzhang@hust.edu.cn}

\author{Bang Wang}
\affiliation{%
  \institution{Huazhong University of Science and Technology (HUST)}
  \city{Wuhan}
  \country{China}
}
\email{wangbang@hust.edu.cn}

\renewcommand{\shortauthors}{Jiazheng Zhang and Bang Wang}

\begin{abstract}
Can we employ one neural model to efficiently dismantle many complex yet unique networks? This article provides an affirmative answer. Diverse real-world systems can be abstracted as complex networks each consisting of many functional nodes and edges. Percolation theory has indicated that removing only a few vital nodes can cause the collapse of whole network. However, finding the least number of such vital nodes is a rather challenging task for large networks due to its NP-hardness. Previous studies have proposed many centrality measures and heuristic algorithms to tackle this network dismantling (ND) problem. Different from theirs, this article tries to approach the ND task by designing a neural model which can be trained from tiny synthetic networks but will be applied for various real-world networks. It seems a discouraging mission at first sight, as network sizes and topologies are quite different across distinct real-world networks. Nonetheless, this article initiates insightful efforts of designing and training a \textit{neural influence ranking model} (NIRM). Experiments on fifteen real-world networks validate its effectiveness for its mostly requiring fewer vital nodes to dismantle a network, compared with the state-of-the-art competitors. The key to its success lies in that our NIRM can efficiently encode both local structural and global topological signals for ranking nodes, in addition to our innovative labelling method in training dataset construction.
\end{abstract}



\maketitle

\section{Introduction}
\label{sec:Intro}
Most real world systems, like airport transportation~\cite{Colizza:et.al:2006:PNAS,Verma:et.al:2014:SciReports}, power grid \cite{Albert:et.al:2004:PhyRevE,Cuadra:et.al:2015:Energies}, Internet infrastructure~\cite{Yook:et.al:2002:PNAS} and etc., can be abstracted as complex networks each consisting of many functional nodes and edges, denoted by $\mathcal{G} = (\mathcal{V},\mathcal{E})$ with $N$ nodes in $\mathcal{V}$ and $M$ edges in $\mathcal{E}$. In practical applications, it has been often reported that the failure of a few nodes and/or edges could significantly degrade the functionality and stability of a network, even leading to the system collapse. A typical example was the Italian nation-wide blackout on September 28 2003 due to cascading failures of power stations starting from a single powerline~\cite{Crucitti:et.al:2004:PhysicalA}. Another recent example was the outage of the Internet service of the American operator Century Link~\cite{Goodwin:Jazmin:2020:CNN} on August 30, 2020 arising from a single router malfunction.

\par
Many efforts have been devoted on studying the dynamics and properties of network stability under different failure scenarios~\cite{Saberi:2015:PhysRep, Almeira:et.al:2020:PhyRevE}. Some insights have been gained from studying network failure processes, including cascading failure~\cite{Buldyrev:et.al:2010:Nature}, network percolation and phase transition~\cite{Callaway:et.al:2000:Phy.Rev.L, Karrer:et.al:2014:Phy.Rev.L, Morone:et.al:2015:Nature}. One of important insights is that a single node can disenable its attached edge(s), which may further cause the failure of its neighboring nodes. Furthermore, if a few nodes are disenabled at the same time, their removals could not only cause local structural damages but also lead to global topological collapse. This observation has recently motivated lots of studies on how to select a few \textit{vital nodes} to attack, such that their removals can dismantle a network, i.e., the network dismantling problem.

\par
The \textit{network dismantling} (ND) problem is to determine a \textit{target attack node set} (TAS), denoted by $\mathcal{V}_t$ ($\mathcal{V}_t \subseteq \mathcal{V}$), for its removal leading to the disintegration of a network into many disconnected components, among which the \textit{giant connected component} (GCC), denoted as $\mathcal{G}\backslash \mathcal{V}_t$, is smaller than a predefined threshold. The objective of network dismantling is to find such a TAS $\mathcal{V}^* \subseteq \mathcal{V}$ with the minimum size, that is,
\begin{equation}\label{Eq:NDObjective}
\mathcal{V}^* \equiv \min \big \{\mathcal{V}_{t} \subseteq \mathcal{V} : | \mathcal{G} \backslash \mathcal{V}_{t} |/N \leq \Theta \big \}
\end{equation}
where $\Theta$ is the predefined dismantling threshold. The ND problem has been proven as NP-hard~\cite{braunstein:et.al:2016:PNAS}, and many solutions have been proposed to find a suboptimal TAS for large networks~\cite{Mugisha:Zhou:2016:PhyRevE, Zdeborov:et.al:2016:SciReports, Fan:et.al:2020:NatureML, Grassia:et.al:2021:NatureC}.

\par
Recently, some researchers have considered on applying deep learning techniques for problems of combinatorial optimization on graph~\cite{Dai:et.al:2017:NIPS, Quentin:et.al:2021:IJCAI, schuetz:et.al:2022:NatureML}. For example, many \textit{Graph Neural Networks} (GNNs)~\cite{Hamilton:et.al:2017:NIPS,Wu:et.al:2020:TNLS,Zhou:et.al:2020:AIOpen} have been designed and applied to find feasible solutions for a large class of graph optimization problems, like the maximum cut, minimum vertex cover, maximum independent set problem.
Also some GNNs have been designed to approximate some global centralities of a node~\cite{Cyrus:et.al:2021:KDD, Changjun:et.al:2019:CIKM, Sunil:et.al:2019:CIKM}, which normally with high computation complexities due to graph-wide operations. For example, the betweenness centrality requires to first find the shortest path for any pair of nodes. With the powerful representation capability of a GNN, such graph centralities or metrics can be approximated with high accuracy and low computational complexity.

\par
For network dismantling, it is also very intriguing to ask the question about whether we can design and train a neural model to output the target attack node set for any input network? However, this question has not yet been well studied in the literature to our best knowledge. This may be due to the fact that not only network sizes but also topological characteristics are quite different across different real-world networks, which seems to discourage applying one neural model to dismantle different networks. Nonetheless, this article presents an affirmative answer to the question through our initiative efforts of designing an effective neural model for network dismantling.

\par
In this article, we are interested in the following question: \textit{Can we find a smallest TAS to dismantle any real-world network via a neural model?} That is, given an input graph $\mathcal{G}$ with its adjacency matrix $\mathbf{A}$, a neural model can output a \textit{dismantling score} $s_{i}^{dis} \in \mathbb{R}$ for each node $v_i$ to facilitate the selection of target attack nodes. In this article, we design and train a neural model, called \textit{neural influence ranking model} (NIRM), for the network dismantling problem. Some challenges on designing and training a neural model have been addressed in this article:
\begin{itemize}
\item Design a neural model: The influence of a node to the stability of a whole network should be not only evaluated from its own structural characteristics, but also compared with other farther apart nodes.
\item Train a neural model:  The \textit{training dataset} as well as the so-called \textit{ground truth} labels for training a neural model are not available, not even mention to their appropriateness and trustfulness.
\end{itemize}
In NIRM, we learn both local structural and global topological characteristics for each node, so as to compute and fuse its local and global influence scores for outputting each node a dismantling score. Our NIRM model is trained by some tiny synthetic networks of no more than thirty nodes, where we can find the optimal TAS(s) via exhaustive search. We design a labelling rule for selected target nodes and propose a training score propagation algorithm to obtain labels for other nodes. We conduct extensive experiments on various real-world network, and results validate the effectiveness of our neural model: Compared with the classic approaches and state-of-the-art competitors, the proposed NIRM performs the best for most of real-world networks.

\par
The rest of the paper is organized as follows: Section~\ref{sec:Related} reviews the related work. The design and training details of our NIRM are introduced in Section~\ref{sec:NIRM} and Section~\ref{sec:Training}, respectively. Section~\ref{sec:Experiment} presents the experimental results. Finally, Section~\ref{sec:Conclusion} concludes the paper.

\begin{figure*}[t]
\centering
\includegraphics[scale = 1.5,width=\textwidth]{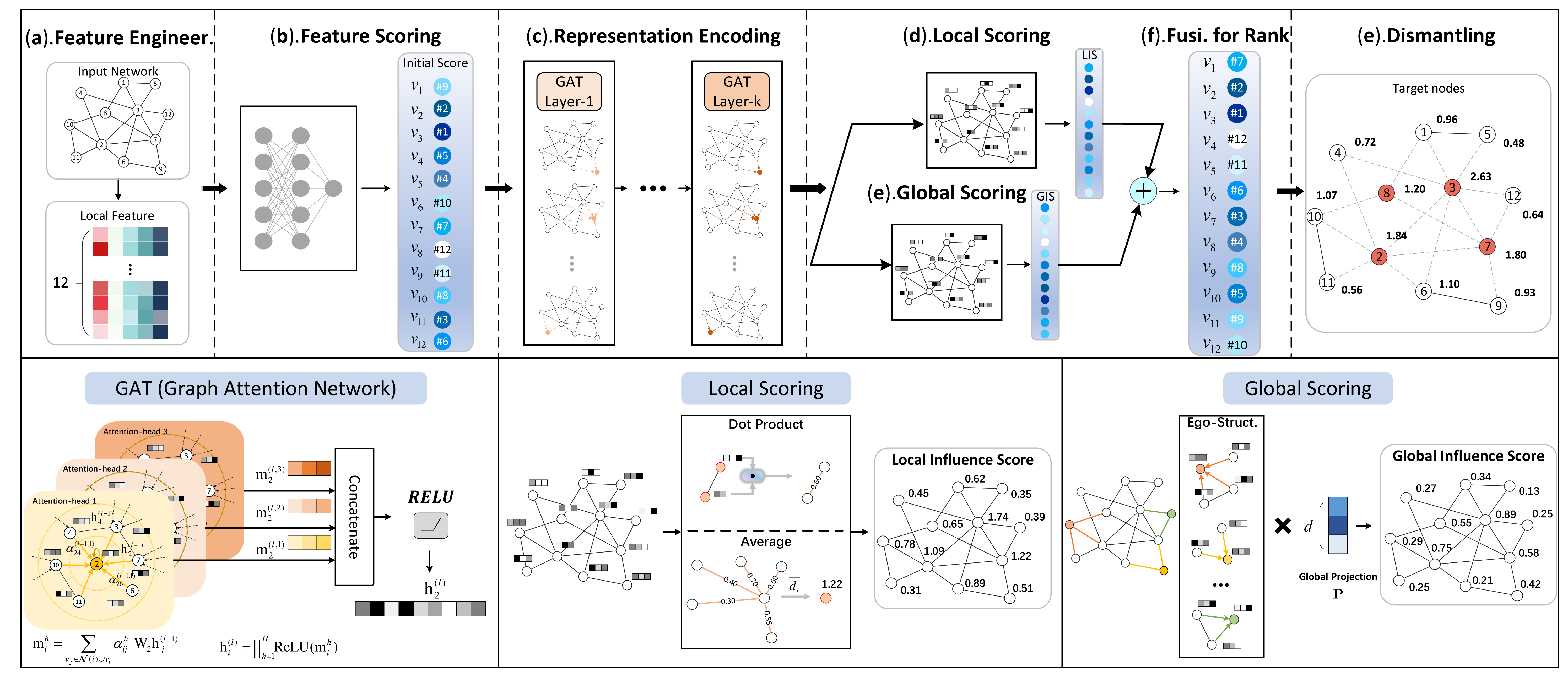}
\caption{The NIRM framework: The upper left part is feature scoring that generates initial scores by converting local features. The lower left part illustrates GAT, which encodes initial scores between neighbors as well as neighborhood structure for learning each node representation. Local scoring and global scoring respectively evaluates a node's influence to the stability of node-centric local structure and network-wide global topology. The final dismantling score is a fusion of local score and global score. The upper right part marks the selected attack nodes in red, which are the same as one of optimal solutions by exhaustive search.}
\label{Fig:NIRM}
\end{figure*}
\begin{figure*}[h]
\centering
\includegraphics[ width=\textwidth, scale =2 ]{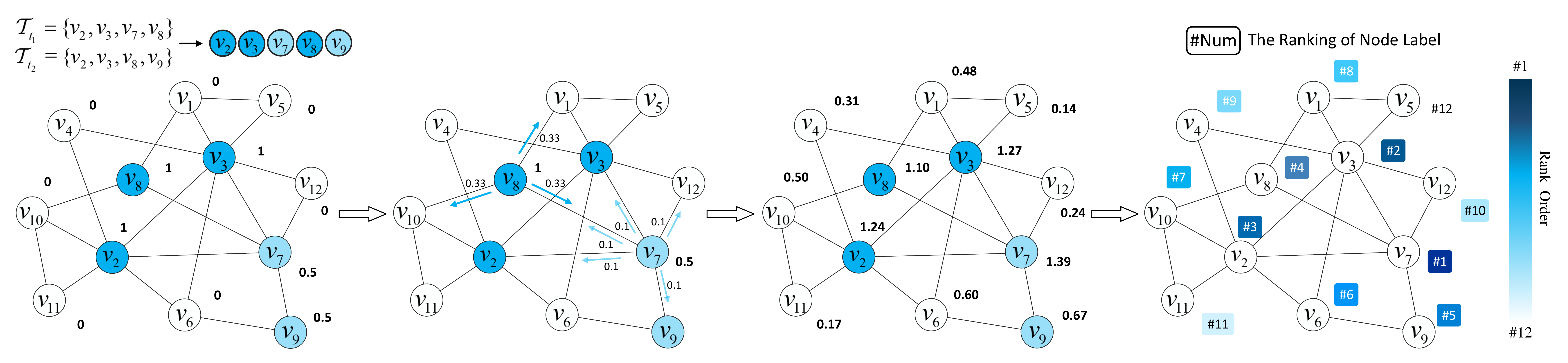}
\caption{Illustration of training score initialization, propagation and labeling. For a tiny synthetic network, exhaustively search the optimal dismantling sets ($\{v_2, v_3, v_7, v_8\} $ and $\{v_2, v_3, v_8, v_9\}$ in the given example), and initialize the training score for a node as its normalized appearances in the optimal dismantling sets. For each vital node (colored blue),  its initial score is equally propagated to its one-hop neighbors. After score propagation, a node training score is the sum of its own and its received scores (if any), of which the normalized rank is ground truth influence label.
The number next to each node in the third figure is its training score. The forth figure presents the ranking results according to the training scores. }
\label{Fig:ScorePropagation}
\end{figure*}

\section{Related work}
\label{sec:Related}

\subsection{Deep Learning on Node Ranking}
Recently, the problem of ranking nodes on a graph has been revisited from deep learning viewpoint and some neural models have been designed. 

\par
Centrality-oriented approaches focus on fast approximating the relative rank of nodes in terms of their centralities to reduce computation complexity~\cite{Grando:et.al:2018:CSUR, Changjun:et.al:2019:CIKM, Sunil:et.al:2019:CIKM, Appan:Mahardhika:2021:IEEEBD, Matheus:et.al:2021:TNSE, Sunil:et.al:2021:TKDD}. Grando et al.~\cite{Grando:et.al:2018:CSUR} propose to estimate eight graph centralities by a neural network with degree and eigenvector as input features. Fan et al.~\cite{Changjun:et.al:2019:CIKM} propose a GNN-based encoder-decoder framework to select the top-K highest betweenness centralities. Sunil et al.~\cite{Sunil:et.al:2021:TKDD} propose a neural model to predict both betweenness and closeness centrality.

\par
Some other neural ranking approaches do not estimate nodes' centralities, but directly output ranking scores for some downstream tasks~\cite{SongQi:et.al:2018:CIKM, Namyong:et.al:2019:KDD, Euyu:et.al:2020:Scientific,Yaojing:et.al:2020:CIKM}. For example, Song et al.~\cite{SongQi:et.al:2018:CIKM} introduce a variant of Recurrent Neural Network for node ranking in heterogeneous temporal graphs to dynamically estimate temporal and structural nodes' influences over time. Yu et al.~\cite{Euyu:et.al:2020:Scientific} propose a CNN-based model to identify critical nodes in temporal networks. Park et al.~\cite{Namyong:et.al:2019:KDD} presented an attentive GNN for predicate-aware score aggregation to estimate entity importance in Knowledge Graphs.

\subsection{Network dismantling}
Network dismantling is a typical discrete combinatorial optimization problem. Attacking different TAS will lead to inestimable combinatorial effects on network connectivity.
Most existing solutions to the ND problem can be generally divided into two categories: centrality metric-based and network decycling-based. For the former, the basic idea is to first compute some centrality metric for each node, and then rank nodes and select the top-$K$ important nodes to form TAS. Some commonly used centralities include degree centrality (DC)~\cite{Albert:et.al:2000:Nature}, closeness centrality (CC)~\cite{Bavelas:Alex:1950:JASA}, betweeness centrality (BC)~\cite{Freeman:Linton:1977:Sociometry}, eigenvector centrality (EC)~\cite{Bonacich:et.al:1987:AmericaJS} and etc. Recently,
Collective Influence (CI)~\cite{Morone:et.al:2015:Nature} was proposed as an improved version of DC, which quantifies the importance of a node by considering not only its one-hop neighbors but also its high-order neighbors.

\par
The network decyling-based approaches, including the Minsum~\cite{braunstein:et.al:2016:PNAS}, BPD~\cite{Mugisha:Zhou:2016:PhyRevE} , CoreHD~\cite{Zdeborov:et.al:2016:SciReports} and etc.,  tried to first finding those nodes whose removals can cause an acyclic network, that is, a network does not contain loops. Next, a kind of greedy tree-breaking algorithms are used to break the residual forest into small disconnected components. In addition, after determining the TAS, some nodes can be reinserted back into the original network, if such reinsertion would not cause the increase of the residual GCC. For example, the BPD~\cite{Mugisha:Zhou:2016:PhyRevE} applies the spin glass theory to solve the feedback vertex set problem when searching nodes for breaking all network loops. Besides, the GND~\cite{Ren:et.al:2018:PNAS} proposes a spectral approach unifying the equal graph partitioning and vertex cover problem, iteratively partitioning the network into two components.

\section{NIRM: A Neural Influence Ranking Model for Network Dismantling}
\label{sec:NIRM}
The proposed neural influence ranking model (NIRM) takes the adjacency matrix $\mathbf{A}\in \mathbb{R}^{N\times N}$ of a network as input, and outputs a vector of \textit{dismantling scores} $\mathbf{s} \in \mathbb{R}^{N\times 1}$ for all nodes. Although different input networks are with different sizes (viz., different $N$), we design several neural modules not only for converting network dimensions but also for learning nodes' representations. Note that all learnable parameters in neural modules are trained from tiny synthetic networks. We compute local and global influence scores based on learned nodes' representations and fuse them to output dismantling score.

\par
Fig.~\ref{Fig:NIRM} presents the NIRM architecture, which consists of the following modules: (1) feature engineering, (2) feature scoring, (3) representation encoding, (4) local scoring, (5) global scoring, and (6) fusion and rank.

\subsection{Feature engineering} This module is to construct a feature vector for each node based on the adjacency matrix of input network. Although many attributes and measures, either local node-centric or global network-wise, can be used as features, we prefer to focus on those local ones, as they do not incur too much computation burden. Specifically, our NIRM uses the following five local node-centric attributes to construct a \textit{local feature matrix} $\mathbf{X}\in \mathbb{R}^{N \times 5}$, where each row is the feature vector $\mathbf{x}_i$ for node $v_i$, consisting of (1) the number of its one-hop neighbors, viz., its degree; (2) the number of its two-hop neighbors; (3) average degree of its one-hop neighbors; (4) its local clustering coefficient; (5) a constant for regulation. These features contains the basic neighborhood information of a node, revealing its local connectivity and structure characteristics to some extent.

\subsection{Feature scoring} This module is to make an initial evaluation of  the importance to network stability for each node based on its local feature. On the one hand, we want to measure a node's connectivity and structure characteristics; On the other hand, we also would like to make a network-wide importance estimation from local feature across all nodes. To this end, we use a fully connected neural network with a network-wide shared kernel to convert local features into scores:
\begin{equation}\label{Eq:FeatureScoring}
s_i^{init} = \operatorname{ReLU} ( \mathbf{W}_1 \mathbf{x}_{i}^{\mathsf{T}} + \mathbf{b}_1),
\end{equation}
where $s_i^{init}$ is the initial score of node $v_i$, $\mathbf{W}_1$ and $\mathbf{b}_1$ are the shared kernel with learnable parameters. We note that other neural networks can also be used for initial scoring.

\subsection{Representation encoding} This module is to encode the initial score of each node together with its neighbors' scores, yet in a discriminate way, into a representation vector in a latent space. After feature engineering and scoring, the initial score could reflect a node's influence to network stability to some extent.
However, the initial score obtained from network-wide conversion only focuses on statistical metrics in the node-centric ego-net, which may not well exploit neighbors' properties as well as multi-relational interactions between neighbors.
We would like to further mine potential neighborhood interactions for evaluating the influence of a node specific to its structure. To this end, we employ the graph neural network technique for encoding nodes' representations.


\par
We introduce Graph Attention Network (GAT)~\cite{Petar:et.al:2018:ICLR} consisting of $L$ \textit{neighborhood aggregation} (NA) layers to learn nodes' representations from their initial scores and high-order structure information. The input is the initial score vector $\mathbf{s}^{init}$, and output is the representation matrix $\mathbf{H}$. Each NA layer applies $H$ \textit{attention heads} to encode node-centric structural properties from different views. We take a node $v_i$ for example to introduce the core operations in $l$-th NA layer:
\begin{itemize}
\item Compute the coefficients $\alpha_{i j}^h$ between $v_i$ and its one-hop neighbors $v_j \in \mathcal{N}_{i}$ (including itself) by the $h$-th attention head $\mathbf{a}_{h}$ ($\mathbf{a}_h^{\mathsf{T}}$ as its transposition):
    \begin{equation}
        \alpha_{i j}^{h} = \frac{\exp [\operatorname{LeakyReLU}(\mathbf{a}_h^{\mathsf{T}}\mathbf{h}_{ij})]}{\sum_{v_k \in \mathcal{N}_i \cup v_i} \exp [\operatorname{LeakyReLU}(\mathbf{a}_h^{\mathsf{T}}\mathbf{h}_{ik})]},
    \end{equation}
    where $\mathbf{h}_{ij}=\mathbf{h}_i^{(l-1)} || \mathbf{h}_j^{(l-1)}$ stands for the concatenation of the hidden embeddings from the $(l-1)$-th NA layer.

\item Aggregate the converted neighborhood hidden embeddings in a weighted way to obtain an aggregation embedding $\mathbf{m}_i^h$ for the $h$-th attention head:
    \begin{equation}
        \mathbf{m}_i^h = \sum_{v_j \in \mathcal{N}(i) \cup v_i} \alpha_{i j}^h \mathbf{W}_2 \mathbf{h}^{(l-1)}_{j}.
    \end{equation}

\item Concatenate the converted aggregation embeddings for $H$ attention heads to output the hidden embedding $\mathbf{h}_i^{l}$ for the $l$-th NA layer.
    \begin{equation}
    \mathbf{h}_i^{(l)}  = \|_{h=1}^{H} \operatorname{ReLU}(\mathbf{m}_i^h).
    \end{equation}
\end{itemize}
Neighboring nodes influence each other through their own initial scores as well as local structural properties. For one NA layer, low-order structural information are attentively encoded for node representation learning. Through multiple NA layers, $L$-hop apart neighbors are also included in representation learning, which can help to capture high-order neighbors' influences as well as some high-order topological information.

\subsection{Local scoring} This module is to evaluate the local influence of each node, for its removal, to the destruction of the local network structure centered to the node. From a node-centric view, one could expect that if a node is with more similar neighbors, then its removal may not only influence the connectivity of its local structure, but also cause further instability cascade through its similar neighbors. For such considerations, we propose to compute a local influence score $s_i^{local}$ for each node $v_i$ as follows:
\begin{equation}\label{Eq:LocalScore}
s_{i}^{local} = \frac{1}{|\mathcal{N}_i|} \sum_{v_j \in \mathcal{N}_i} \langle \mathbf{h}_{i}, \mathbf{h}_{j} \rangle + \bar{d}_i,
\end{equation}
where $\langle \cdot, \cdot \rangle$ stands for the dot product of two vectors, and $\bar{d}_i = d_i/d_{max}$ is the normalized node degree and $d_{max}$ the largest degree of input network.

\subsection{Global scoring} This module is to compare the global influence of one node, for its removal, to the possible damage of the global topology with that of other nodes. Consider that two nodes are with similar local structure but distant from each other. As the dismantling is for the whole network, it is necessary to further distinguish the importance of the two nodes for their global ranking. For such considerations, we design a global projection operator $\mathbf{P}$ to learn a global influence score for each node:
\begin{equation}\label{Eq:GlobalScore}
s_i^{global} = \langle \mathbf{P}^{\mathsf{T}}, \hat{\mathbf{h}}_i \rangle,
\; \mathrm{where} \;  \hat{\mathbf{h}}_i = \sum_{v_j \in \mathcal{N}_{i} \cup v_i} \frac{1}{\sqrt{|\mathcal{N}_i|+1}} \cdot \mathbf{h}_j,
\end{equation}
where $\mathbf{P}$ denotes the learnable projection vertor and $\hat{\mathbf{h}}_i$ is the neighborhood representation of node $v_i$. We note that using $\hat{\mathbf{h}}_i$ instead of $\mathbf{h}_i$ is again to enjoy node-centric local structural information such that global comparison is conducted for a kind of \textit{ego-structure} though represented by a single node.

\subsection{Fusion and rank} This module is to fuse the local score $s_i^{local}$ and global score $s_i^{global}$ as the \textit{dismantling score} $s_i^{dis}$ for each nodes. As we have already designed sophisticated mechanisms for computing $s_i^{local}$ and $s_i^{global}$, we simply add up the two scores for  $s_i^{dis}$, that is,
\begin{equation}\label{Eq:DismantlingScore}
s_i^{dis} =  s_i^{global} +  s_i^{local}.
\end{equation}
Finally, nodes are ranked according to their dismantling scores.

\subsection{Complexity analysis}
The NIRM model requires node features matrix $\mathbf{X}$ and the sparse adjacency matrix $\mathbf{A}$ as input, which can be applied to any given network to infer dismantling scores. The time complexity of NIRM consists of three parts, the first part explicitly learns the initial score from local feature, which takes $O(\left| V \right|)$ and $V$ is the number of nodes. The second part GAT applies the self-attention mechanism to encode high-order structure and neighbors information into topology representation, the time complexity of the embedding process is $O\left( L \left( \left| V \right| + \left| E \right| \right) \right) $, where $L$ is the number of propagation layers(e.g., 3), $E$ is the number of edges. After GAT, then combine the local influence scores and global influence scores to sort nodes in the entire graph, where the time complexity of calculating the local and global influence scores is $O(\left| E \right|)$, $O(\left| V \right|)$ respectively and the node ranking operation takes $O(\left| V \right| log \left| V \right|)$. Therefore, the overall complexity of the NIRM model is given by $ O( \left| V \right| +  \left| E \right| + \left| V \right| log \left| V \right|) $.

\begin{table}[t]
\centering
\renewcommand{\arraystretch}{1.2}
\setlength{\tabcolsep}{0.6mm}
\caption{Statistical properties of real-world networks}
\label{tab:SI_Dataset}
\begin{tabular}{l c c c c c c }
\hline
Network & Category & Nodes & Edges & Density & AVD & Diameter \\
\hline
PPI~\cite{Dongbo:et.al:2003:NAR}                                    & Protein          & 2,224     & 6,609      & 0.0027         & 5.94           & 11    \\
HI-II-14~\cite{Thomas:et.al:2014:Cell}       & Protein          & 4,165     & 13,087     & 0.0016         & 6.28           & 11    \\
Ca-GrQc~\cite{Leskovec:Krevl:2014:SNAP}        & Collab.    & 4,158     & 13,422    & 0.0016          & 6.46           & 17  \\
NetScience~\cite{Kunegis:et.al:2013:www}                            & Collab.    & 1,461     & 2,742     & 0.0026          & 3.75           & 17  \\
DNCEmails~\cite{Kunegis:et.al:2013:www}                             & Comm.    & 1,866     & 4,384      & 0.0025         & 4.70           & 8   \\
Innovation~\cite{Kunegis:et.al:2013:www}                            & Comm.    & 241       & 923        & 0.0319         & 7.66           & 5   \\
Infectious~\cite{Kunegis:et.al:2013:www}                            & Contact          & 410       & 2,765      & 0.0330         & 13.49          & 9     \\
Genefusion~\cite{Hoglund:et.al:2006:Oncogene}                       & Gene             & 291       & 279        & 0.0066         & 1.92           & 9     \\
P-H~\cite{Kunegis:et.al:2013:www}                                   & Social           & 2,000     & 16,098    & 0.0081          & 16.10          & 10   \\
HM~\cite{Kunegis:et.al:2013:www}                                    & Social           & 1,858     & 12,534    & 0.0073          & 13.49          & 14   \\
UsPower~\cite{Watts:et.al:1998:Nature}                              & Grid        & 4,941     & 6,594      & 0.0005         & 2.67           & 46    \\
Crime~\cite{Kunegis:et.al:2013:www}                                 & Malicious        & 829       & 1,473     & 0.0043          & 3.55           & 10   \\
Corruption~\cite{Ribeiro:et.al:2018:JCN}                            & Malicious        & 309       & 3,281     & 0.0689          & 21.24          & 7     \\
Roget~\cite{Vladimir:Andrej:2006:Pajek}      & Lexicon          & 1,010     & 3,646     & 0.0072          & 7.22           & 10  \\
Bible~\cite{Kunegis:et.al:2013:www}                                 & Lexicon          & 1,773     & 9,131     & 0.0058          & 10.30          & 8   \\
\hline
\end{tabular}
\end{table}

\begin{table}[t]
\centering
\renewcommand{\arraystretch}{1.2}
\caption{NIRM model configurations and hypermeters.}
\begin{tabular}{l|c}
\hline
Hyper-parameter & Value  \\
\hline
Adam optimizer learning rate           & $1 \times{10^{-3}}$ \\
regularization term (L2 penalty)       & $1 \times{10^{-4}}$ \\
learning decay                         & $0.4$           \\
mini-batch size                        & $6$            \\
number of self-attention heads         & $8, 4, 2$       \\
maximum training epoches               & $50$            \\
neighbor-aggregation layers            & $3$             \\
dimensions of node embedding           & $32, 16, 8$     \\
patience period                        & $8$              \\
probability of dropout                 & $0.1, 0.2$       \\
negative slope of Leaky ReLU           & $0.2$       \\
number of neurons per layer            & $5, 8$          \\
\hline
\end{tabular}
\label{Table:Hyperparameter}
\end{table}

\begin{figure*}[t]
\centering
\includegraphics[width = 17.8cm]{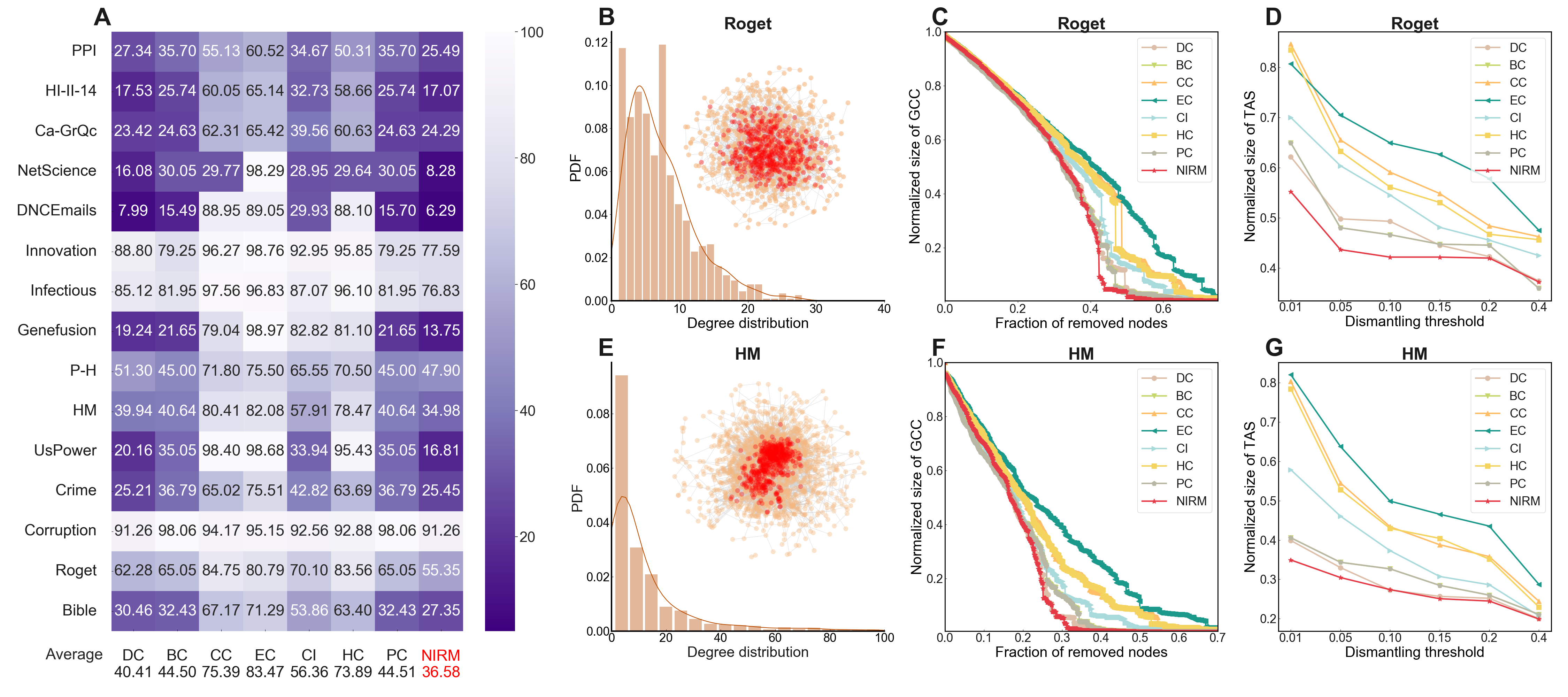}
\caption{Comparison of one-pass dismantling performance on real-world networks.
(\textit{A}): The smaller the normalized size of TAS $\rho$, the darker the color. Among 15 real world networks, the NIRM achieves the best in 12 networks, and second best on 3 networks. The bottom row provides the averaged $\rho$ over all 15 networks, where ours is 36.58 and second best is 40.41. (\textit{B} and \textit{E}): the node degree distribution and the attack nodes (red) selected by NIRM for two real world networks: Roget containing 1,010 nodes and 3,646 edges, and HM containing 1,858 nodes and 12,534 edges. (\textit{C} and \textit{F}) : the NGCC when removing different fractions of target attack nodes, where the area of ours is 298.79 and 296.06, the second smallest is BC of 303.28 in Roget,DC of 302.55 in HM. (\textit{D} and \textit{G}): the dismantling performance $\rho$ against different dismantling thresholds $\Theta$.}
\label{fig:Real_Onepass_Experiment}
\end{figure*}

\begin{figure*}[h]
\centering
\includegraphics[width = 17.8cm]{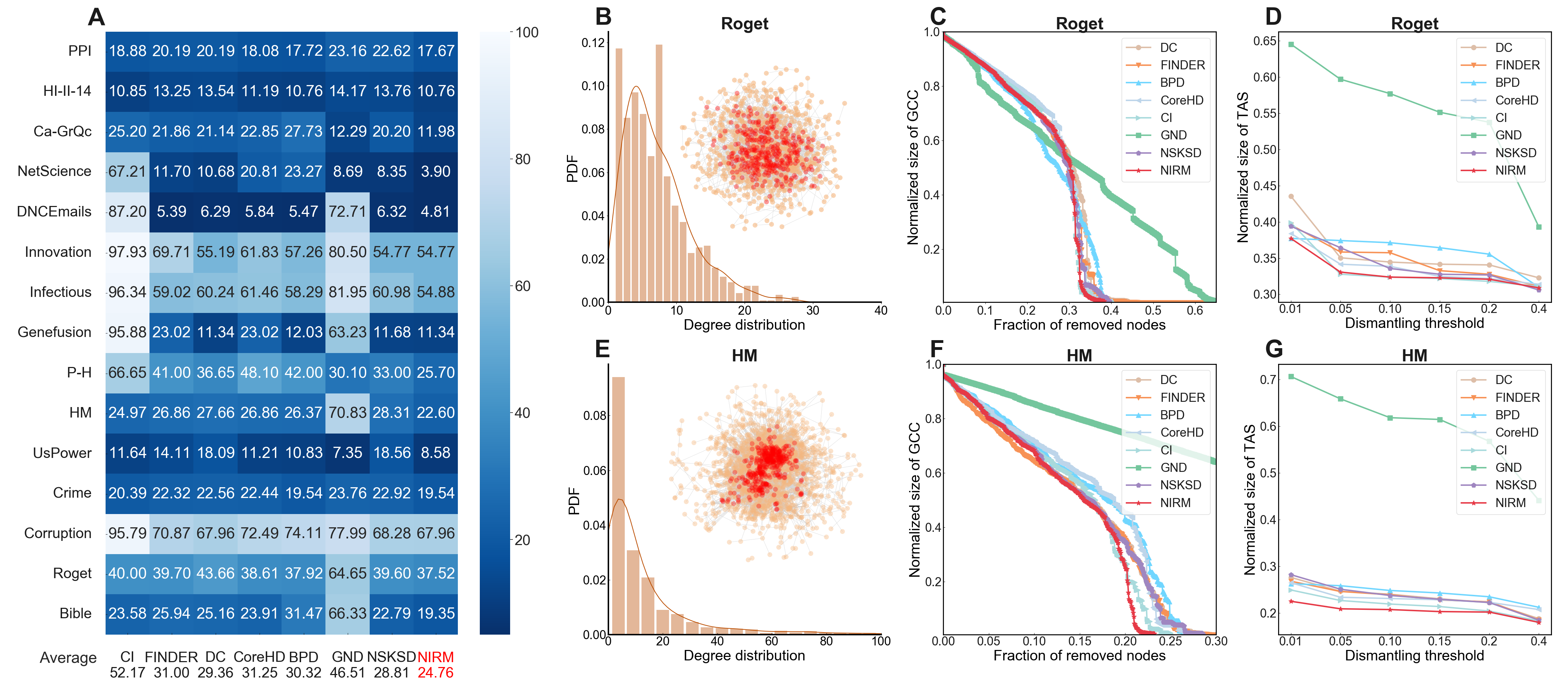}
\caption{Comparison of adaptive dismantling performance on the fifteen real-world networks. (\textit{A}) The smaller the normalized size of TAS $\rho$, the darker the color. Among the 15 networks, the NIRM achieves the best on 14 network, and the second best on 1 networks. The bottom row provides the average $\rho$, where ours is 24.76 and the second best is 28.81. (\textit{B} and \textit{E}): The selected attack nodes (red) by NIRM are fewer than those of one-pass dismantling: 37.52$\%$ vs. 55.35$\%$ in Roget, and 22.60$\%$ vs. 34.98$\%$ in HM. (\textit{C} and \textit{F}): the NGCC when removing different fractions of target attack nodes. (\textit{D} and \textit{G}): the dismantling performance $\rho$ against different dismantling thresholds $\Theta$.
}
\label{fig:Real_Repeated_Experiment}
\end{figure*}

\section{Model Training}
\label{sec:Training}

\subsection{Training datasets} We first introduce how to construct training datasets from synthetic model networks. In network science, some generative models have been widely accepted for producing synthetic networks, including Erd\"{o}s-R\'{e}nyi (ER, $p=0.1$) \cite{Erdos:et.al:1960:evolution}, Watt-Strogatz (WS, $k=4, p=0.1$) \cite{Watts:et.al:1998:Nature}, Barab\'{a}si-Albert (BA, $m=3$) \cite{Barabasi:et.al:1999:Science} and Powerlaw-Cluster (PLC, $m=3, p=0.05$) \cite{Holme:et.al:2002:PhyRevE}. Such synthetic model networks can well reflect one or more kinds of topological characteristics of many real-world networks, such as uniform or power-law node degree distribution. We use these generative models to produce a large number of tiny synthetic networks to compose a training dataset. Each synthetic network contains only a few of nodes (randomly selected from 20 to 30).

\par
We use \textit{exhaustive search} to first find the optimal TAS(s) for each tiny synthetic network. Note that the computational cost of exhaustive search increases exponentially with the increase of network size. This is also why we only use small-scale synthetic networks for training dataset. Further notice that for a synthetic training network $\mathcal{G}_t=(\mathcal{V}_t, \mathcal{E}_t)$, there could exist more than one optimal TAS. For $\mathcal{G}_t$, let $\mathcal{T}_t$ denote the set of its optimal TASs. For each node $v_i \in \mathcal{T}_t$, we count its frequency of appearing in different TASs, denote by $N_i$. Furthermore, let $N_{max}$ denote the maximum of $N_i$. Then for each $v_i \in \mathcal{T}_t$, we set its initial training score $c_i^0 = N_i / N_{max}$.

\par
Although the aforementioned approach can set a so-called ground-truth score for each node $v_i \in \mathcal{T}_t$, there could exist some node $v_j \in \mathcal{V}_t \backslash \mathcal{T}_t$ with no such a score, i.e., $c_j^0=0$. On the one hand, such a node $v_j$ may still play a role to network stability, though only with a smaller influence. On the other hand, the objective of our neural model is to output dismantling scores for final ranking. To take care of all nodes and avoid the unbalance of score, we propose the following \textit{training score propagation} algorithm to propagate the initial training score of a node $v_i \in \mathcal{T}_t$ to its one-hop neighbors:
\begin{equation}\label{Eq:ScorePropagation}
c_i =  \sum_{v_j \in \mathcal{N}_i} \frac{c_j^0}{|\mathcal{N}_j|}  + c_i^0, \; \forall c_i \in \mathcal{V}_t,
\end{equation}
where $\mathcal{N}_j$ is the set of $v_j$'s neighbors and $c_j^0$ denotes initial training score of $v_j$, which is equally allocated to $v_j$'s neighbors (divided by the number of its neighbors). After score propagation, training score $c_i$ of node $v_i$ is the sum of its own and its received scores.

\par
Fig.~\ref{Fig:ScorePropagation} illustrates the training score initialization, propagation and labeling by an example network.

\subsection{Loss function}
Let $\mathbf{c}$ denote the nodes' scores of an instance in the training dataset, and $\hat{\mathbf{s}}$ is the estimated scores of NIRM. The loss of one training network is defined as the following \textit{mean square error}:
\begin{equation}\label{Eq:Loss}
Loss = \frac{1}{|\mathcal{V}|} \sum_{v_i \in \mathcal{V}} (c_i - \hat{s}_i)^2.
\end{equation}
We implemented the NIRM in the PyTorch framework and used the Adam optimizer to train the model. During the training, we decay the learning rate with an early stopping criterion based on the loss on the validation set.

\par
In our NIRM, the learnable parameters include $\mathbf{W}_1$, $\mathbf{b}_1$ in feature scoring, $\mathbf{a}_{l,h} (l=1,...,L, h=1,...,H)$, $\mathbf{W}_2$ in representation learning, and $\mathbf{P}$ in global scoring. There are in total $854$ learnable parameters in our model. We generate 4,000 synthetic model networks with $95\%$ as training instances and $5\%$ as validation instances.

\section{Experiment Results and Analysis}
\label{sec:Experiment}

\subsection{Experiment settings}
We evaluate NIRM on fifteen real-world networks from various domains, including infrastructure network, communication network, collaboration network, malicious network and etc. Table \ref{tab:SI_Dataset} presents the characteristics and statistics of these real-world networks, model configurations and training hype-parameters can be found in Table \ref{Table:Hyperparameter}. All the experiments were conducted on an 8-core workstation with the following configurations: Intel Xeon E5-2620 v4 CPU with 2.10GHz, 32GB of RAM and Nvidia GeForce GTX 1080Ti GPU with 11GB memory. We release our code on Github at: https://github.com/JiazhengZhang/NIRM.

\subsection{Dismantling strategy}
For an input network, our NIRM outputs all nodes' dismantling scores for ranking, so enabling two dismantling strategies. The first one, called \textit{one-pass dismantling}, just selects the top-$K$ nodes according to their dismantling scores, such that the removal of only such $K$ nodes can lead to the \textit{normalized size of GCC} (NGCC) in the residual network not exceeding the threshold $\Theta$. NGCC is defined as the number of nodes in GCC divided by that of the original network. The second one, called \textit{adaptive dismantling}, only selects the top-$1$ node to remove; While the residual network after removing such one selected node will be input again to output a new node for next removal; Such selection-and-removal process will be repeated until the NGCC satisfies the threshold.

\subsection{Basline methods}
We compare NIRM with other state-of-the-art approaches which focus on estimating node importance, approaches for network dismantling can be generally classified into two categories according to the corresponding dismantling strategies.

\textbf{One-pass approaches}, common metric-based approaches mostly belong to this category, various baselines are selected and their introduction are listed below:
\begin{itemize}
\item DC (Degree Centrality)~\cite{Albert:et.al:2000:Nature}. Sequentially remove nodes in order of degree centrality.

\item CI (Collective Influence)~\cite{Morone:et.al:2015:Nature}. CI is defined as the product of the node degree (minus one) and total degrees of neighboring nodes at the surface of a ball of constant radius.

\item EC (Eigenvector Centrality)~\cite{Bonacich:et.al:1987:AmericaJS}. EC is based on the idea that the more important a neighboring node is connected, the more important that node is.

\item BC (Betweeness Centrality)~\cite{Freeman:Linton:1977:Sociometry}. BC is a path-based metric that counts the number of shortest paths through node.

\item CC (Closeness Centrality)~\cite{Bavelas:Alex:1950:JASA}. CC measures the average distance between the node and other nodes.

\item HC (Harmonic Centrality)~\cite{Paolo:et.al:2014:IM}. HC is a variant of the CC algorithm that can be applied to disconnected network.

\item PC (Percolation Centrality)~\cite{Piraveenan:et.al:2013:PloS}. PC quantifies relative impact of nodes based on their topological connectivity, as well as their percolation states.
\end{itemize}

\begin{figure*}[t]
\centering
\includegraphics[width=\textwidth, height= 0.45\textwidth]{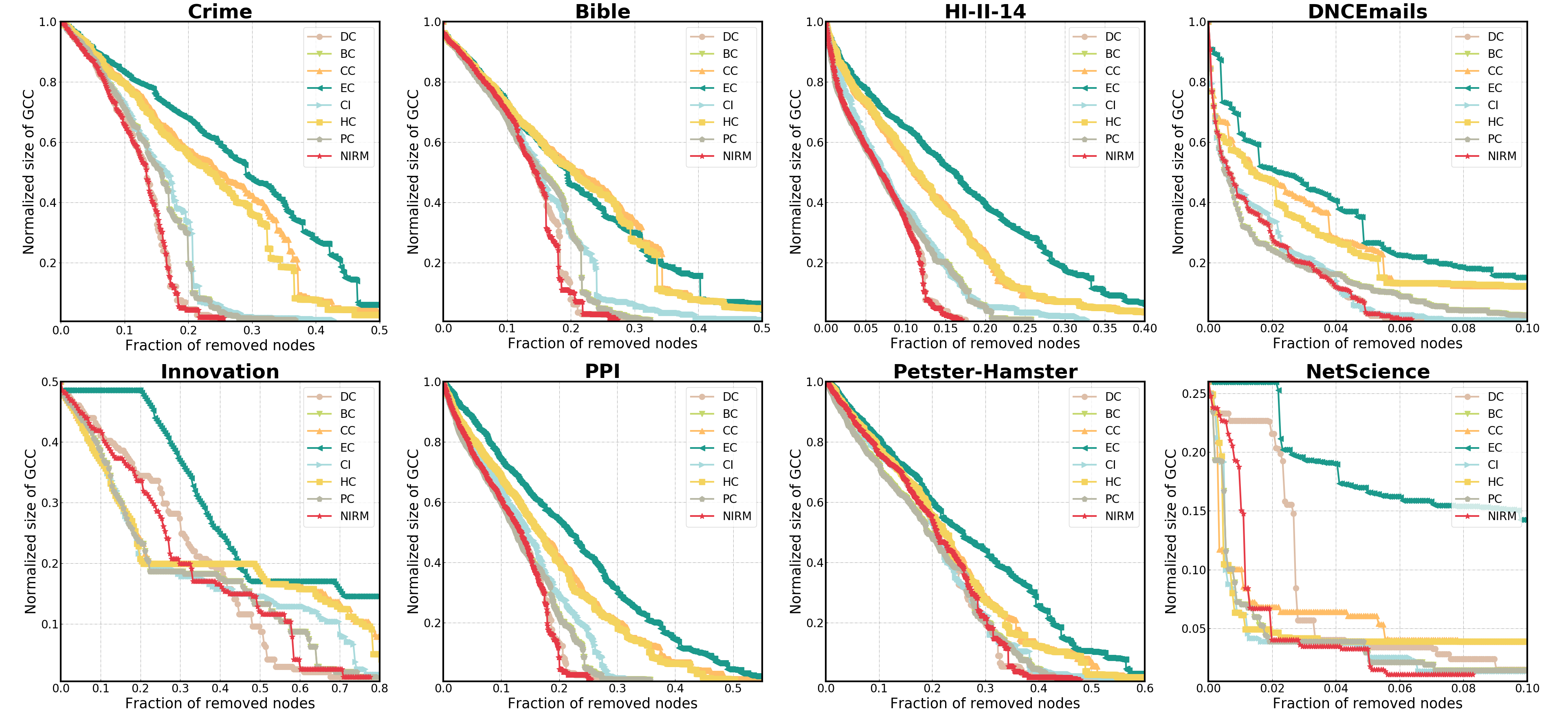}
\caption{The NGCC when removing different fractions of target attack nodes
under one-pass dismantling.}
\label{fig:SI_Real_Onepass}
\end{figure*}

\begin{figure*}[ht]
\centering
\includegraphics[width=\textwidth, height= 0.45\textwidth]{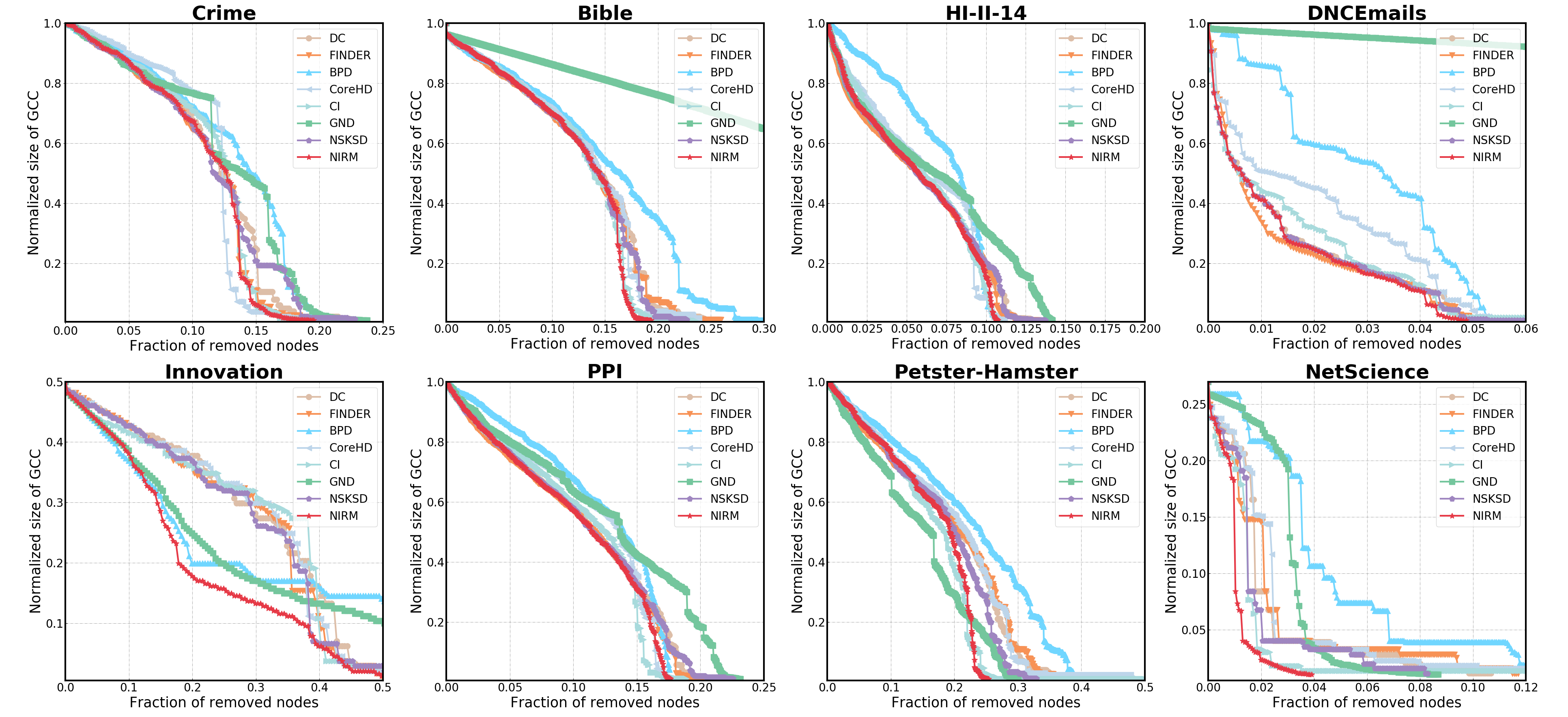}
\caption{The NGCC when removing different fractions of target attack nodes
under adaptive dismantling.}
\label{fig:SI_Real_Repeated}
\end{figure*}

\par
\textbf{Adaptive approaches}, we also explore the performance of adaptive approaches on network dismantling, details of these approaches are described as follows:
\begin{itemize}
\item DC~\cite{Albert:et.al:2000:Nature}. The adaptive version of DC, which recalculates the node degree centrality in the remaining network after each removal.

\item CI~\cite{Morone:et.al:2015:Nature}. The adaptive version of CI, which measure the importance of nodes based on the local structure of a ball of radius around every node.

\item BPD~\cite{Mugisha:Zhou:2016:PhyRevE}. BPD considers the spin glass theory and proposes a belief propagation-guided decimation algorithm. It first remove nodes until eliminating all the loops in the residual network. After that, BPD check the size of tree components and iteratively removes the root node.

\item CoreHD~\cite{Zdeborov:et.al:2016:SciReports}. CoreHD is similar with BPD in that both have eliminating loops and tree breaking. First remove the nodes with the highest degree from the 2-core of the network in an adaptive way. After the 2-core is empty, it adopts greedy tree-breaking algorithm.

\item GND~\cite{Ren:et.al:2018:PNAS}. GND proposes a spectral approach of node-weighted Laplacian operator. In each step, it partitions the remaining network into two components, and remove a set of nodes which covers all of the edges between two non-overlapping components at minimal cost.

\item FINDER~\cite{Fan:et.al:2020:NatureML}. FINDER is a reinforcement learning-based method. It encodes the state (remaining network) and all possible actions (remove node) into embedding vectors, so that it take action $a$ that represents the maximum expected rewards given state $s$.

\item NSKSD~\cite{Yiguang:et.al:2021:TCS}. NSKSD consider the overlapping effects between nodes, and introduced novel metrics KSD and NS. It proposed the double-turns selection mechanism to remove nodes in an adaptive way.
\end{itemize}
Note that some centrality-based approaches, e.g. BC, CC, EC, are not included in adaptive approaches, though they can be applied for adaptive dismantling, their adaptive versions are with great computation burdens yet not satisfactory dismantling performance. In addition, we implement NIRM as well as other state-of-the-art approaches without node reinsertion.

\begin{table}[t]
\centering
\caption{Ablation study of dismantling performance $\rho$ when using node intermediate scores and final scores to select attack nodes.}
\small
\renewcommand{\arraystretch}{1.2}
\setlength{\tabcolsep}{1.6mm}
(\textbf{A}) Adaptive dismantling
\begin{tabular}{l| c c c c c c}
\hline
Methods & UsPower & P-H & Ca-GrQc & Infectious & Bible  & HM \\
\hline
\hline
NIRM-IS      & 12.37   & 53.95   & 32.23 & 65.61  & 41.06 & 26.43 \\
NIRM-GS      & 67.86   & 89.35   & 74.43 & 96.83  & 83.42 & 89.24 \\
NIRM-LS      & 9.05    & 27.55   & 13.32 & 55.85  & 20.47 & 24.17 \\
NIRM         & \textbf{8.58}    & \textbf{25.70}   & \textbf{11.98} & \textbf{54.88}  & \textbf{19.35} & \textbf{22.60} \\
\hline
\end{tabular}

\vspace{10pt}

(\textbf{B}) One-pass dismantling
\begin{tabular}{l| c c c c c c}
\hline
Methods & UsPower  & P-H  & Ca-GrQc  & Infectious  & Bible   & HM \\
\hline
\hline
NIRM-IS      & 18.94    & 86.80   & 35.95 & 79.02  & 31.75 & 48.44 \\
NIRM-GS      & 98.91    & 96.45   & 97.55 & 98.78  & 94.47 & 97.42 \\
NIRM-LS      & 18.32    & 49.15   & 24.63 & 81.71  & 28.26 & 38.64 \\
NIRM         & \textbf{16.81}   & \textbf{47.90}   & \textbf{24.29}  & \textbf{76.83}  &  \textbf{27.35}  & \textbf{34.98} \\
\hline
\end{tabular}
\label{Table:Ablation}
\end{table}

\subsection{Performance evaluation}
We first compare the one-pass dismantling performance of our NIRM with that of seven commonly used ranking metrics. Fig.~\ref{fig:Real_Onepass_Experiment}~\textit{A} presents the dismantling performance in terms of the \textit{normalized size of TAS} (denoted by $\rho$, and $\rho=|\mathcal{V}_{tas}|/|\mathcal{V}|$) for fifteen real-world networks, when setting the dismantling threshold $\Theta = 0.01$. Fig.~\ref{fig:Real_Onepass_Experiment}~\textit{B} and ~\textit{E} visualize the attack nodes (red) in Roget and HM to illustrate that our model is capable of identifying critical nodes for network dismantling. For different dismantling thresholds $\Theta$ in Roget and HM, as shown in Fig.~\ref{fig:Real_Onepass_Experiment}~\textit{D} and \textit{G}, NIRM consistently outperforms competitors. Our NIRM achieves the smallest $\rho$ in twelve real-world networks, and three smallest in three networks. On average, the NIRM selects about $36.58\%$ attack nodes to dismantle a network; While the second best one requires attacking $40.41\%$ nodes.

\par
We next focus on the adaptive dismantling strategies. In general, as each iteration of selection-and-removal is for a particular input network, it can be expected that the performance of adaptive algorithm would be better than its one-pass version. We note that although the BC, CC, HC, PC and EC algorithm can also be applied for adaptive dismantling, they are not included for comparison, as their adaptive versions are with great computation burdens yet with not satisfactory dismantling performance. On the other hand, we include seven state-of-the-art schemes for performance comparison, including DC~\cite{Albert:et.al:2000:Nature}, CI~\cite{Morone:et.al:2015:Nature}, BPD~\cite{Mugisha:Zhou:2016:PhyRevE}, CoreHD~\cite{Zdeborov:et.al:2016:SciReports}, GND~\cite{Ren:et.al:2018:PNAS}, FINDER~\cite{Fan:et.al:2020:NatureML} and NSKSD~\cite{Yiguang:et.al:2021:TCS}.

\par
Fig.~\ref{fig:Real_Repeated_Experiment}~\textit{A} compares the adaptive dismantling performance for the same fifteen real-world networks. It is not unexpected that the adaptive versions of NIRM, DC, and CI requires fewer attack nodes to meet the same dismantling threshold. Among the fifteen real-world networks, our NIRM achieves the best in fourteen ones, and second best in one. Furthermore, on average, our NIRM requires to attack only $24.76\%$ nodes, with an improvement of $4.05\%$ compared with the second best NSKSD requiring $28.81\%$ attack nodes. Fig.~\ref{fig:Real_Repeated_Experiment}~\textit{C} and \textit{F} plot the NGCC when removing different fractions of target attack nodes. We observe that the NIRM can often achieve the smallest NGCC against the same number of removed nodes, indicating our model better captures the response of system throughout the whole attack process. In addition, the area under such a curve can evaluate the average attack efficiency of a dismantling scheme. The areas of our NIRM are $248.27$, and $250.84$ in HM, and Roget, respectively, much smaller than other current popular methods FINDER of $259.81$, CI of $253.56$ in the corresponding network.
See Fig.~\ref{fig:SI_Real_Onepass} and Fig.~\ref{fig:SI_Real_Repeated} for more dismantling results on eight real-world networks.

\par
Table~\ref{Table:Ablation} reports the results of our ablation study, where six real-networks are used to examine the normalized size of TAS performance of adaptive and one-pass dismantling. Recall that in our NIRM, three intermediate scoring vectors, i.e., initial scores, local scores and global scores, can also be extracted for ranking nodes and dismantling networks. It can be seen that although local scores can already achieve competitive dismantling performance, its fusion with global scores further reduces target attack nodes. This collaborates our design objective of using a global projection to further make up network-wide comparisons of nodes' influences.

\section{Conclusion}
\label{sec:Conclusion}
Although many heuristic algorithms have been proposed for tackling the network dismantling problem, little has been done on employing recent deep learning techniques. The main challenge is from the fact that real world networks are with diverse sizes and distinct characteristics, which seems to discourage the use of a single neural model for different networks. In addition, training a neural model faces the difficulties of unknown ground truth labels, especially for large networks. Nonetheless, this article has provided an insightful trial of designing and training a neural influence ranking model for network dismantling. The key design philosophy is to encode local structural and global topological characteristics of each node for ranking its influence to network stability. Another factor of success is from our score propagation for only recruiting tiny synthetic networks for model training.

\par
Experiments on fifteen real-world networks have validated our NIRM as a promising solution to the network dismantling task. Yet we would like to note again that it is this style of training from tiny but applying for real-world networks, could open a new window for further investigations. Indeed, we acknowledge that our neural model is far from perfection, and many advances can be expected. In NIRM, only a few local attributes are crafted as initial features. A possible extension is to also take care of the node-centric motifs~\cite{Shao:et.al:2021:TKDD} for feature encoding. Also our global project kernels are somewhat plain, while sophisticated modules like Parametric UMAP~\cite{Sainburg:et.al:2021:NeuralCompute} and techniques like contrastive learning~\cite{You:et.al:2020:NIPS} can also be tried. Finally, we would like to expect further investigations on more delicate labeling strategies as well as training mechanisms.

\begin{acks}
This work is supported in part by National Natural Science Foundation of China (Grant No: 62172167).
We also want to use our NIRM model on MindSpore\footnote{http://www.mindspore.cn/}, which is a new deep learning computing framework. These problems are left for future work.
\end{acks}

\bibliographystyle{ACM-Reference-Format}
\balance
\bibliography{main_paper}




\end{document}